\documentclass[prb, twocolumn,superscriptaddress,preprintnumbers,a4paper,amsmath,amssymb,showpacs,floatfix]{revtex4}

\usepackage{graphicx}
\usepackage{color}\color[rgb]{0.000,0.000,0.000} 
\usepackage{amssymb} 
\usepackage{amsmath} 

\usepackage{dcolumn}
\usepackage{bm}

\begin{document}
\newcommand{\TN}{T$_{N}$}
\newcommand{\Tn}{T$_{N}$}
\newcommand{\TNo}{T$_{N1}$}
\newcommand{\TNt}{T$_{N2}$}
\newcommand{\Tno}{T$_{N1}$}
\newcommand{\Tnt}{T$_{N2}$}
\newcommand{\Tp}{T$^{\prime}$}
\newcommand{\dg}{$^{\circ}$}
\newcommand{\zncrs}{ZnCr$_{2}$S$_{4}$}
\newcommand{\zncrse}{ZnCr$_{2}$Se$_{4}$}
\newcommand{\etal}{\textit{et al.}}

\preprint{Draft 1}

\title{Spin-driven Phase Transitions in ZnCr$_2$Se$_4$ and  ZnCr$_2$S$_4$ Probed by High Resolution
Synchrotron X-ray and Neutron Powder Diffraction}









\author{F. Yokaichiya$^1$, A. Krimmel$^2$, V. Tsurkan$^{2,3}$, I. Margiolaki$^4$,
P. Thompson$^4$, H. N. Bordallo$^1$, A. Buchsteiner$^1$, N.
St\"u{\ss}er$^1$, D. N. Argyriou$^1$, A. Loidl$^2$}
\affiliation{$^1$ Helmholtz Center Berlin for Materials and Energy,
Glienicker Str. 100, Berlin D-14109, Germany\footnote{Formally known
as the Hahn-Meitner-Institut.}\\ $^2$ Experimental Physics V, Center
for Electronic Correlations and Magnetism, Augsburg University, D -
86159 Augsburg, Germany\\ $^3$ Institute of Applied Physics, Academy
of Sciences of Moldova, MD2028, Chisinau, R. Moldova\\ $^4$ European
Synchrotron Radiation Facility, BP 220, F-38043 Grenoble Cedex,
France}

\date{\today}

\begin{abstract}
The crystal and magnetic structures of the spinel compounds
ZnCr$_2$S$_4$ and ZnCr$_2$Se$_4$ were investigated by high
resolution powder synchrotron and neutron diffraction.
ZnCr$_2$Se$_4$ exhibits a first order phase transition at
$T_N=21$~K into an incommensurate helical magnetic structure.
Magnetic fluctuations above $T_N$ are coupled to the crystal
lattice as manifested by negative thermal expansion. Both, the
complex magnetic structure and the anomalous structural behavior
can be related to magnetic frustration. Application of an external
magnetic field shifts the ordering temperature and the regime of
negative thermal expansion towards lower temperatures. Thereby,
the spin ordering changes into a conical structure. ZnCr$_2$S$_4$
shows two magnetic transitions at $T_{N1}=15$~K and $T_{N2}=8$~K
that are accompanied by structural phase transitions. The crystal
structure transforms from the cubic spinel-type (space group
$Fd$\={3}$m$) at high temperatures in the paramagnetic state, via
a tetragonally distorted intermediate phase (space group
$I4_1$/$amd$) for $T_{N2} < T < T_{N1}$ into a low temperature
orthorhombic phase (space group $I m m a$) for $T < T_{N2}$. The
cooperative displacement of sulfur ions by exchange striction is
the origin of these structural phase transitions. The low
temperature structure of ZnCr$_2$S$_4$ is identical to the
orthorhombic structure of magnetite below the Verwey transition.
When applying a magnetic field of 5~T the system shows an induced
negative thermal expansion in the intermediate magnetic phase as
observed in ZnCr$_2$Se$_4$.
\end{abstract}



\maketitle

\section{\label{sec:intro}Introduction}

Chromium spinels of stoichiometry ACr$_2$X$_4$ (X=O, S, Se) are
 frustrated magnets. Frustration is distinguished
by the fact that it is impossible to satisfy all pair-wise
interactions simultaneously. The chromium ions of the spinel
structure form a network of corner-sharing tetrahedra known as the
pyrochlore lattice. For nearest neighbor antiferromagnetic (AFM)
exchange, strong geometric frustration (GF) arises due to the
triangular arrangement of the ions. Additional frustration may be
present due to the competition between ferromagnetic (FM) and AFM
interactions. The different exchange interactions critically
depend on the Cr-Cr distance. At small Cr-Cr separation strong
direct AFM exchange dominates. At larger distances, additional
90$^o$ FM Cr-X-Cr and more complex Cr-X-A-X-Cr superexchange paths
contribute. Despite remarkable different exchange interactions
reflected in paramagentic Curie Weiss temperatures varying from
-400 to +200 K, chromium spinels generally reveal intricate AFM
order below a N\'eel temperature in the range of 10 - 20 K. The
broad range of magnetic exchange strength and magnetic ground
states has recently been summarized in a corresponding phase
diagram \cite{Rudolf07}.

The large degeneracy resulting from frustration may be lifted by
any additional interaction which enters in a non-perturbative way.
This is the origin for the large variety of fascinating ground
states that become available for frustrated systems and that
depend on the very details of the interactions involved (e.g. exchange, anisotropy, magnitude of the spin etc.). Spin-liquids
\cite{Canals98}, spin-ice states \cite{Ramirez99,Bramwell01},
clusters or loops of a finite number of spins
\cite{Garcia00,Radaelli02,Lee02}, heavy fermion-like behavior
\cite{Kondo97,Krimmel99}, as well as singlet formation
\cite{Berg03} are typical examples of these exotic ground states.
In particular in chromium and vanadium oxide spinels magnetic
frustration is lifted via a coupling between spin and lattice
degrees of freedom \cite{Rudolf07,Reehuis03}, resulting in a
magnetic \cite{Tschernyshyov02} or spin-driven \cite{Yamashita00}
Jahn-Teller effect. This is a remarkable behavior, since the
3$d^3$ electronic configuration of Cr$^{3+}$ ions corresponds to a
half filled lower $t_{2g}$ triplet. Hence, a spin-only ion with
almost spherical charge distribution and negligible spin-orbit
coupling is expected. Structural distortions at low temperatures
induced by a gain in magnetic energy is thus a novel kind of
magneto-elastic coupling. For ZnCr$_2$O$_4$ and CdCr$_2$S$_4$,
such a spin-lattice coupling has been discussed in terms of a
spin-Peierls like transition \cite{Lee00,Sushkov05,Chung05}.

However, a strong spin-lattice coupling may not necessarily
involve static lattice distortions but may be purely dynamic in
nature. The coupling of magnetic exchange interactions and long
range magnetic order to phonon modes has been put forward many
years ago \cite{Baltensperger68,Baltensperger70,Bruesch72}. Recent
$ab ~initio$ calculations evidenced a magnetic exchange induced
splitting of phonon modes decoupled from static lattice
distortions \cite{Massidda99,Fennie06}. These findings are in
agreement with recent experimental infrared (IR) studies on the
spin-phonon coupling of a large number of chromium spinels
\cite{Rudolf07}.

ZnCr$_2$S$_4$ exhibits two subsequent AFM transitions at
$T_{N1}=15$~K and $T_{N2}=8$~K, as evidenced by measurements of
the magnetization, specific heat, thermal expansion and IR
spectroscopy \cite{Hemberger06}. As outlined above, a strong
spin-phonon coupling induces a significant splitting of phonon
modes at $T_{N1}$ and $T_{N2}$. The corresponding anomalies in the
specific heat and thermal expansion can be suppressed by an external magnetic field, thus evidencing its
magnetic origin \cite{Hemberger06}. The magnetic susceptibility of
ZnCr$_2$S$_4$ reveals a paramagnetic Curie-Weiss temperature close
to 0 K as a result of the competition between FM and AFM exchange
interactions which are almost equal in strength, a situation
termed bond frustration \cite{Hemberger06}.

ZnCr$_2$Se$_4$ orders in a complex AFM structure below $T_N=21$~K
despite strong FM exchange reflected in a large positive value of
the Curie-Weiss temperature $\theta_{CW}=115$~K \cite{Plumier66}.
The magnetic phase transition is of first order, as evidenced by
sharp anomalies in the specific heat and thermal expansion
\cite{Hemberger07}. The magnetic phase transition can be
suppressed by an external magnetic field. ZnCr$_2$Se$_4$ reveals
negative thermal expansion (NTE) below 75~K down to $T_N$ and
extremely large magnetostriction \cite{Hemberger07}. The magnetic
transition is accompanied by small structural distortions with a
lowering of the symmetry from cubic $Fd$\={3}$m$ to tetragonal
$I4_1/amd$ (Ref.~\onlinecite{Kleinberger66}) or orthorhombic
$Fddd$ (Ref.~\onlinecite{Hidaka03}). The magnetic
origin of the structural distortions is most clearly evidenced in
the phonon spectra of ZnCr$_2$Se$_4$. IR spectroscopy revealed a
pronounced splitting of the IR-active phonon modes at $T_N$ that
can be completely suppressed by the application of an external
magnetic field \cite{Rudolf207}.

Ferroelectricity has attracted a lot of interest, in particular
when strongly coupled to magnetic properties in terms of
multiferroic behavior. This phenomenon has been mainly observed in
the perovskites RMnO$_3$ (R=Rare Earth) \cite{Kimura03} and, more
recently, in other materials \cite{IOP}. It occurs in helical
magnets where the spin spiral plane is perpendicular to the
propagation vector. It has also been observed in the spinel
CoCr$_2$O$_4$ with a conical spiral \cite{Yamasaki06}. New studies
have been made in order to explore multiferroic states in other
kinds of helimagnets in which the magnetic propagation wave vector
is perpendicular to spin spiral plane. However, according to the
spin-current model \cite{Katsura05}, or equivalently the inverse
Dzyaloshinskii-Moriya interaction \cite{Sergienko06,Mostovoy06},
this structure by itself cannot produce ferroelectricity. In order
to obtain an electric polarization in such compounds, an external
magnetic field is needed to be applied unparallel to the
propagation wave vector direction, as demonstrated recently for
ZnCr$_2$Se$_4$ \cite{Murakawa08}.

Although these systems have been extensively studied, to the best
of our knowledge, this paper is the first report on the
coexistence of the diffuse magnetic scattering and NTE in
ZnCr$_2$Se$_4$ at zero and applied magnetic field, as well as in
ZnCr$_2$S$_4$ in external magnetic fields. In addition, clear
structural transformations in  ZnCr$_2$S$_4$ from cubic to
tetragonal at $T_{N1}$ and from tetragonal to orthorhombic at
$T_{N2}$ are documented.

The experimental results of the present paper are presented
separately for ZnCr$_2$Se$_4$ (Section~\ref{sec:expres2}) and
ZnCr$_2$S$_4$ (Section~\ref{ZnCrS}). Firstly, we report on
extensive neutron powder diffraction (NPD) studies of
ZnCr$_2$Se$_4$ in zero magnetic field from 2 to 300 K that
revealed the existence of NTE. Moreover, the magnetic structure
was determined without considering any structural distortions in
order to simplify the analysis. Subsequently, we investigated the
effect of an applied magnetic field on the crystallographic and
magnetic structure of ZnCr$_2$Se$_4$. In the next section, we
present high resolution synchrotron x-ray diffraction measurements
of ZnCr$_2$S$_4$ that provide evidence for modifications of the
crystal structure accompanying the magnetic transitions, followed
by neutron diffraction studies in zero field for obtaining the
magnetic structure of the sulfide. Finally, we show results in
external magnetic fields evidencing field induced NTE in the
sulfide compound, similar to what is observed in the selenide.

\section{\label{sec:exp} Experimental}

Polycrystalline samples of ZnCr$_2$S$_4$ and ZnCr$_2$Se$_4$  were
prepared by conventional solid-state synthesis from high purity
elements at 800 $^o$C in evacuated quartz ampoules. To assure
complete chemical reaction and to achieve single phase material
with good homogeneity, the sintering process was repeated several
times with intermediate grinding and pressing of the material.

High resolution NPD data were measured from both samples using the
fine resolution neutron powder diffractometer E9 located at the
Berlin Neutron Scattering Center (BENSC) of the Helmholtz Center
Berlin. The NPD data were measured in zero field and in an applied
vertical magnetic field of 5~T as a function of temperature in the
range 2~K - 300~K. The experiments were performed using neutron
wavelengths of $\lambda=$ 1.797~\AA\ and 2.4~\AA\ respectively with
a resolution (FWHM) of $\Delta d/d \sim 0.2 \%$ \cite{Tobbens01}.
For \zncrs\ additional high intensity NPD data were measured as a
function of temperature on the E6 diffractometer (also at BESNC)
with a wavelength of $\lambda= 2.4 $~\AA. Both x-ray and neutron
diffraction data were analyzed by the Rietveld method employing
the FULLPROF suite of programs \cite{Rodriguez93}.

The NPD data were modeled for all temperatures and fields
employing the cubic spinel structure which is described by
space group $Fd$\={3}$m$ with the Cr atoms at the Wyckoff position
$16d$, $(0.5, 0.5, 0.5)$ and the Zn ions at the Wyckoff position
$8a$, $(0.125, 0.125, 0.125)$. The Se- and S-atoms reside on
Wyckoff position $32e$ $(x,x,x)$ with $x=$0.2594(1) for
ZnCr$_2$Se$_4$ and $x=$0.2583(1) for ZnCr$_2$S$_4$, respectively.
Typical Caglioti resolution parameters that accounted for the
angular dependence of the width of the Bragg reflections were
refined from the Rietveld analysis to be $U=0.0776^o$ ,
$V=-0.1337^o$ and $W=0.1265^o$. For both compounds, magnetic
scattering was observed below $T_N$ and was simultaneously refined
with the crystal structure within our Rietveld analysis. The
corresponding magnetic models are described below.

High resolution x-ray powder diffraction (XPD) data were measured
from ZnCr$_2$S$_4$ at beamline ID31 of the European Synchrotron
Radiation Facility (ESRF), Grenoble \cite{Fitch04}. The sample was
placed in an aluminum container and mounted in a
liquid-helium-cooled cryostat. The measurements were performed in
a temperature range 2 K - 40 K. An incident wavelength of $\lambda
= 0.39950$~\AA ~in combination with a large range of the
scattering angle $-5.964^o < 2\theta < 67.968^o$ allowed
measurements up to 17.8~\AA$^{-1}$ of the scattering vector $Q=4
\pi sin(\theta) / \lambda$. Three different phases have been
refined simultaneously: the main phase ZnCr$_2$S$_4$, aluminum
scattering arising from the sample holder and cryostat and
Cr$_2$S$_3$ as a residual of the sample preparation procedure. The
refinements indicate a mean volume fraction of 2.3$\%$ of the
spurious phase close to sensitivity limit of conventional powder
diffraction. Apart from instrumental parameters (peak shape and
resolution parameters), an overall scale factor, the lattice
constants, the sulfur position and isotropic temperature factor
for each atomic species were refined.

\section{\label{sec:expres2} Neutron powder diffraction of Z\lowercase{n}C\lowercase{r}$_2$S\lowercase{e}$_4$}

\subsection{Zero field NPD measurements of \zncrse}

Within the experimental accuracy, the NPD data of ZnCr$_2$Se$_4$
are consistent with the well know spinel $Fd$\={3}$m$ crystal
structure at all temperatures and fields.  In zero field, several
new magnetic reflections are detected at low diffraction angles
below $T_N= 21$~K, as shown in Fig. \ref{fig:fig1}. These magnetic
reflections can be indexed by a propagation vector $k\sim$ $( 0,
0, \delta )$ with $\delta\sim$0.44.  It is reported that this
magnetic transition is accompanied by a structural transition to
an orthorhombic space group $F d d d$ \cite{Hidaka03,Hemberger07}
with a small orthorhombic distortion of $c/a=0.9999$ at 20 K and
$a\cong b$ \cite{Hidaka03,Hemberger07,Kleinberger66}. Examination
of the present NPD data revealed no evidence for such a transition
as the size of this distortion is below the detection limit set by
the $Q-$resolution of our instrument. For the analysis of the NPD
data we therefore continued to use the cubic spinel structure also
for the low temperature data below $T_N$.

The magnetic scattering below $T_N$ was modelled using the
spherical harmonics approach and considering four Cr-atoms in the
primitive unit cell stacked along the $c-$axis as reported in the
literature \cite{Akimitsu78, Hidaka03}.  We found that the data
were consistent with a model where Cr-spins form ferromagnetic
layers that are coupled antiferromagnetically along the $c-$axis.
The ordering along the $c-$axis is incommensurate (IC) and
naturally leads to a helical magnetic structure propagating along
the unique $c-$ axis as illustrated in Fig.~\ref{fig:ZnCrSe}.
Rietveld refinement of the 2K data results in a screw angle of
42\dg\ and a magnetic moment of the Cr-atoms of 1.90(2) $\mu_B$,
somewhat higher than the value of 1.54~$\mu_B$ reported in
reference \onlinecite{Akimitsu78}. On warming, we find a change of
the incommensurability $\delta$ towards lower values consistent
with previous reports \cite{Akimitsu78,Plumier75}. The screw angle
and the Cr moment varies smoothly with increasing temperature, as
shown in Fig. ~\ref{fig:magmoment}, reaching final values of 39\dg
and 1.4(2) $\mu_B$ at $T_N=21$~K respectively. The jump of the
ordered magnetic moment at $T_N$ indicates a first order phase
transition, in agreement with heat capacity and thermal expansion
measurements \cite{Hemberger07}. The first order nature of the
magnetic transition is probably related to the accompanying
structural modifications.

\begin{figure}[tb!]
\begin{center}
\includegraphics[scale=0.3]{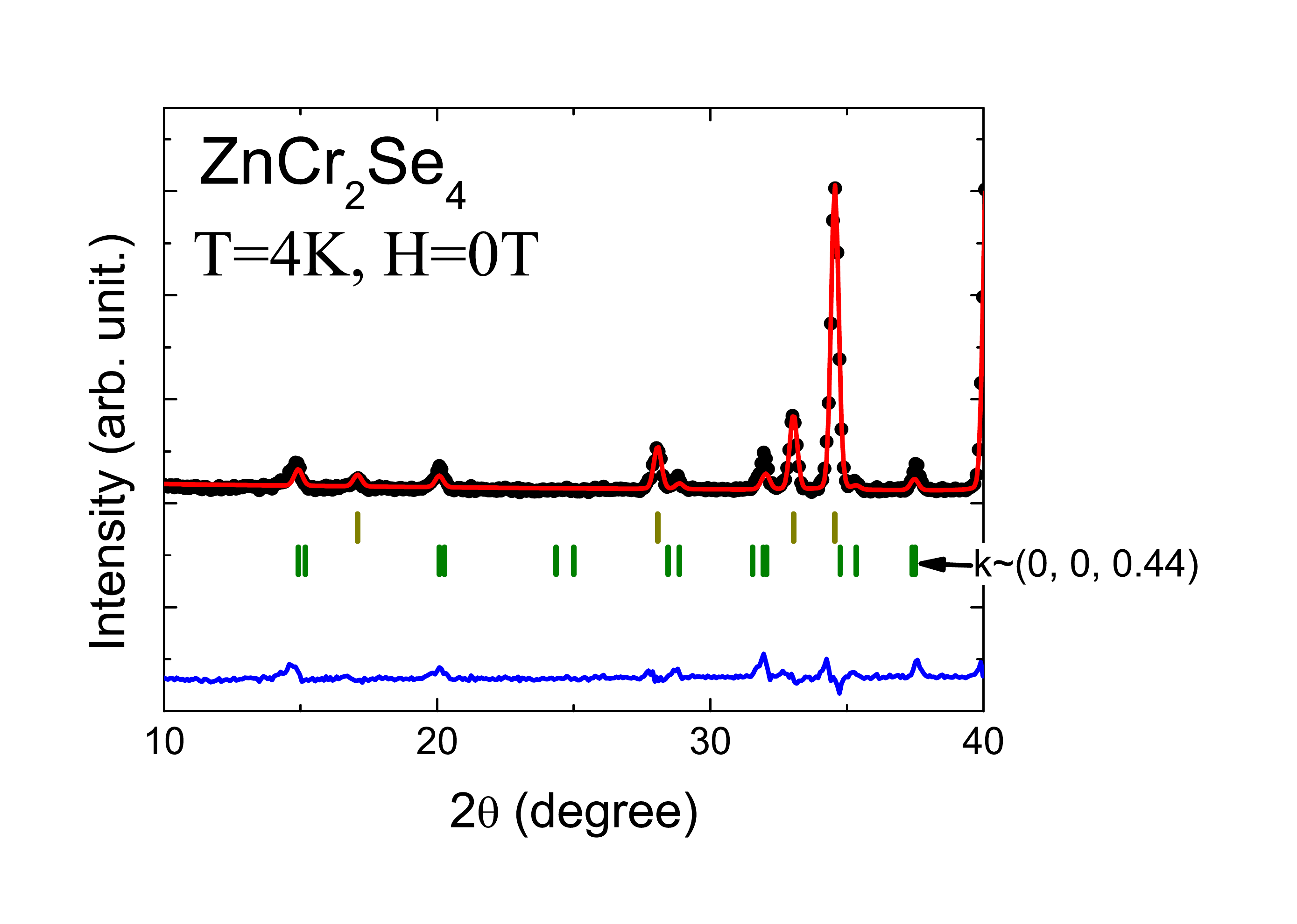}
\caption{(Color online) Rietveld refinement of the NPD date measured from
ZnCr$_2$Se$_4$ at 4~K in zero field. The black points represent
the experimental data while the red and blue lines correspond to
the refinement and the difference between calculated and
experimentally observed intensities, respectively. The upper row of tick
marks shows the calculated positions of nuclear Bragg reflections (dark yellow) and the
lower row indicate the calculated position of magnetic
reflections (dark green) with $k\sim$ $( 0,
0, 0.44 )$ }
\label{fig1}
\end{center}
\end{figure}

\begin{figure}[tb!]
\begin{center}
\includegraphics[scale=0.3]{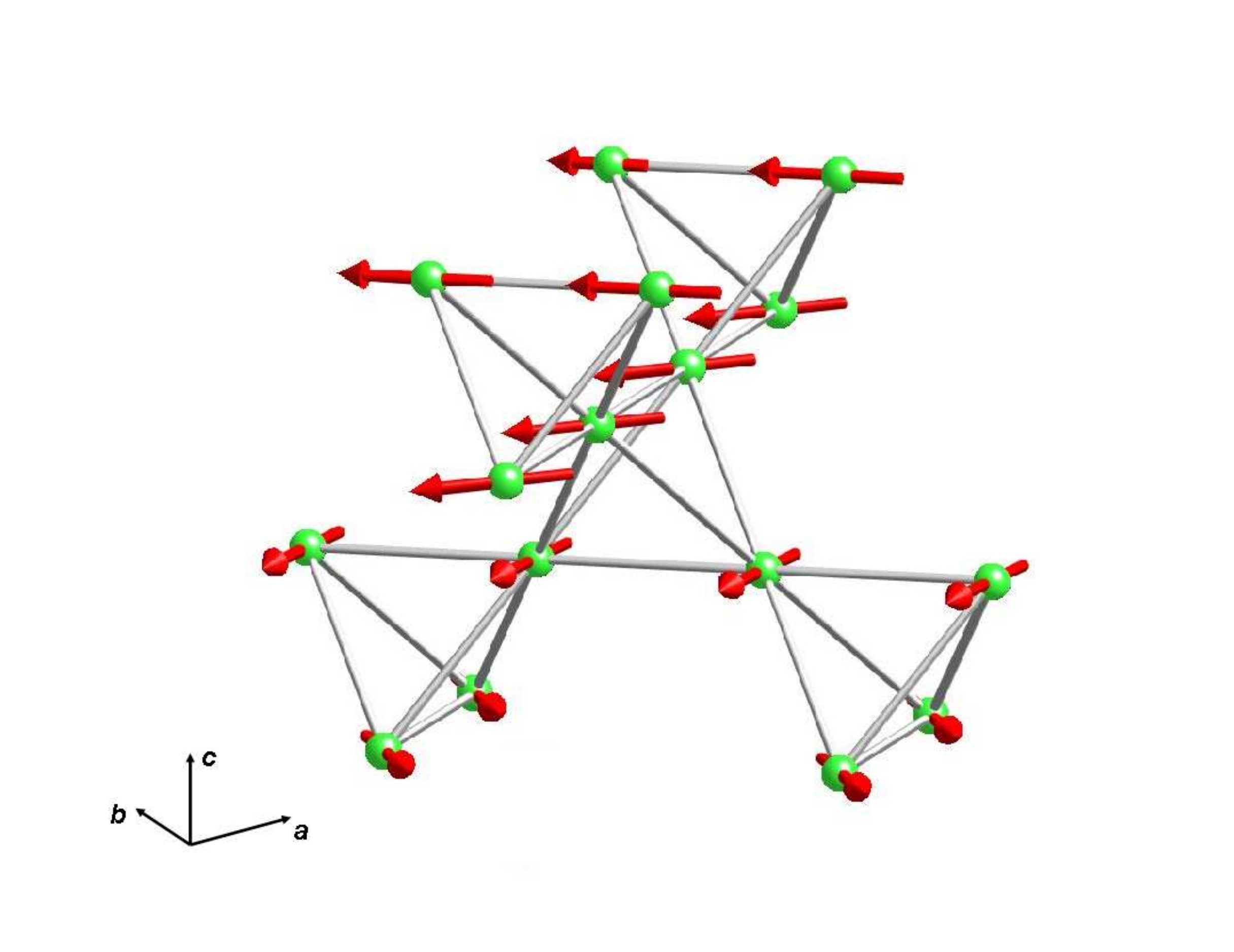}
\caption{(Color online) Schematic representation of the magnetic structure of
ZnCr$_2$Se$_4$. For clarity only the Cr atoms are
displayed. }\label{fig:ZnCrSe}
\end{center}
\end{figure}
\begin{figure}[tb!]
\begin{center}
\includegraphics[scale=0.3]{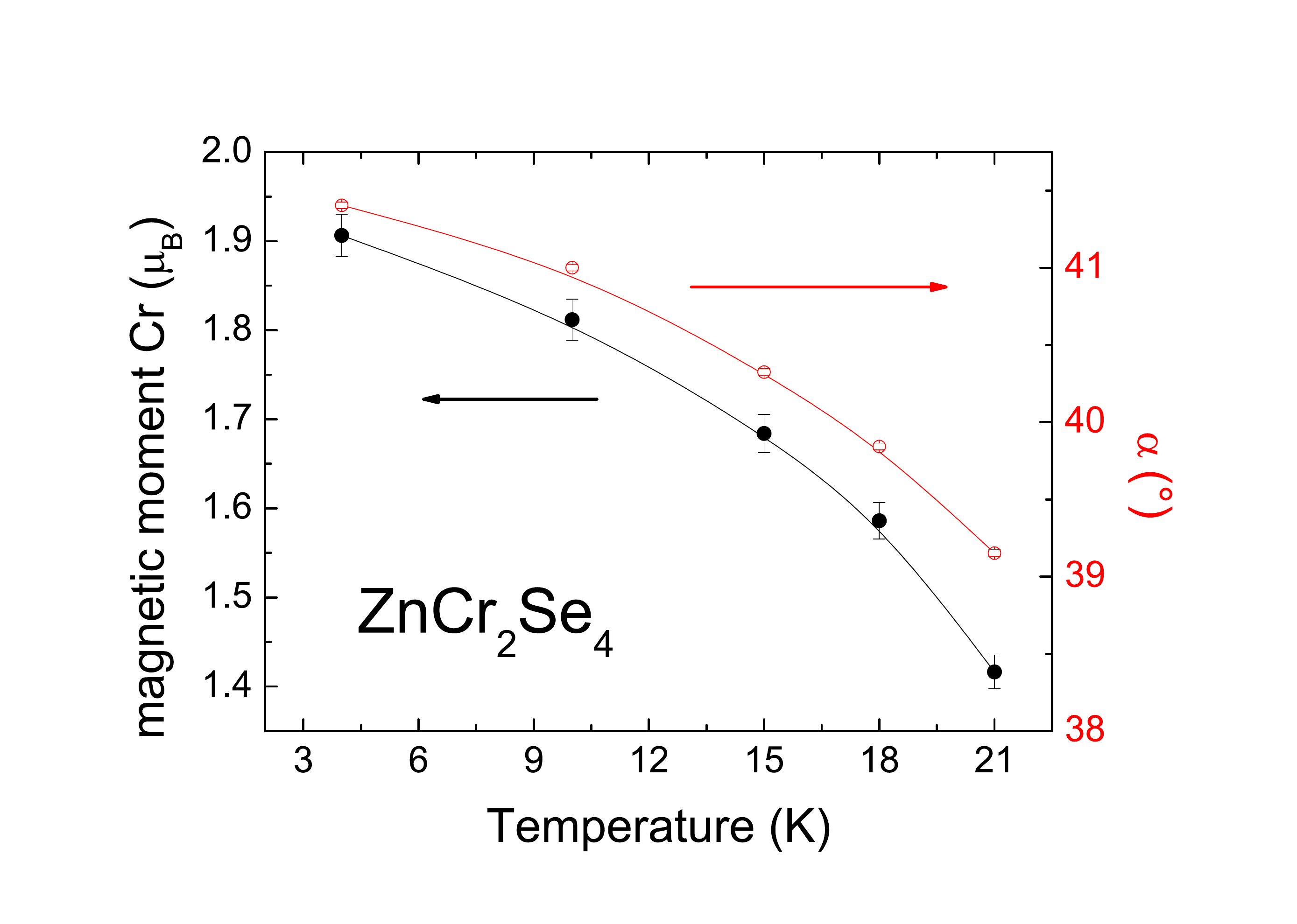}
\caption{\label{fig:magmoment} (Color online) Ordered magnetic
moment of Cr atoms (black, left side axis) and screw angle
$\alpha$ (red, right side axis) vs. temperature in ZnCr$_2$Se$_4$ determined from
Rietveld analysis of the NPD data. Full lines are to guide the eye.}
\end{center}
\end{figure}

\begin{figure}[tb!]
\begin{center}
\includegraphics[scale=0.6]{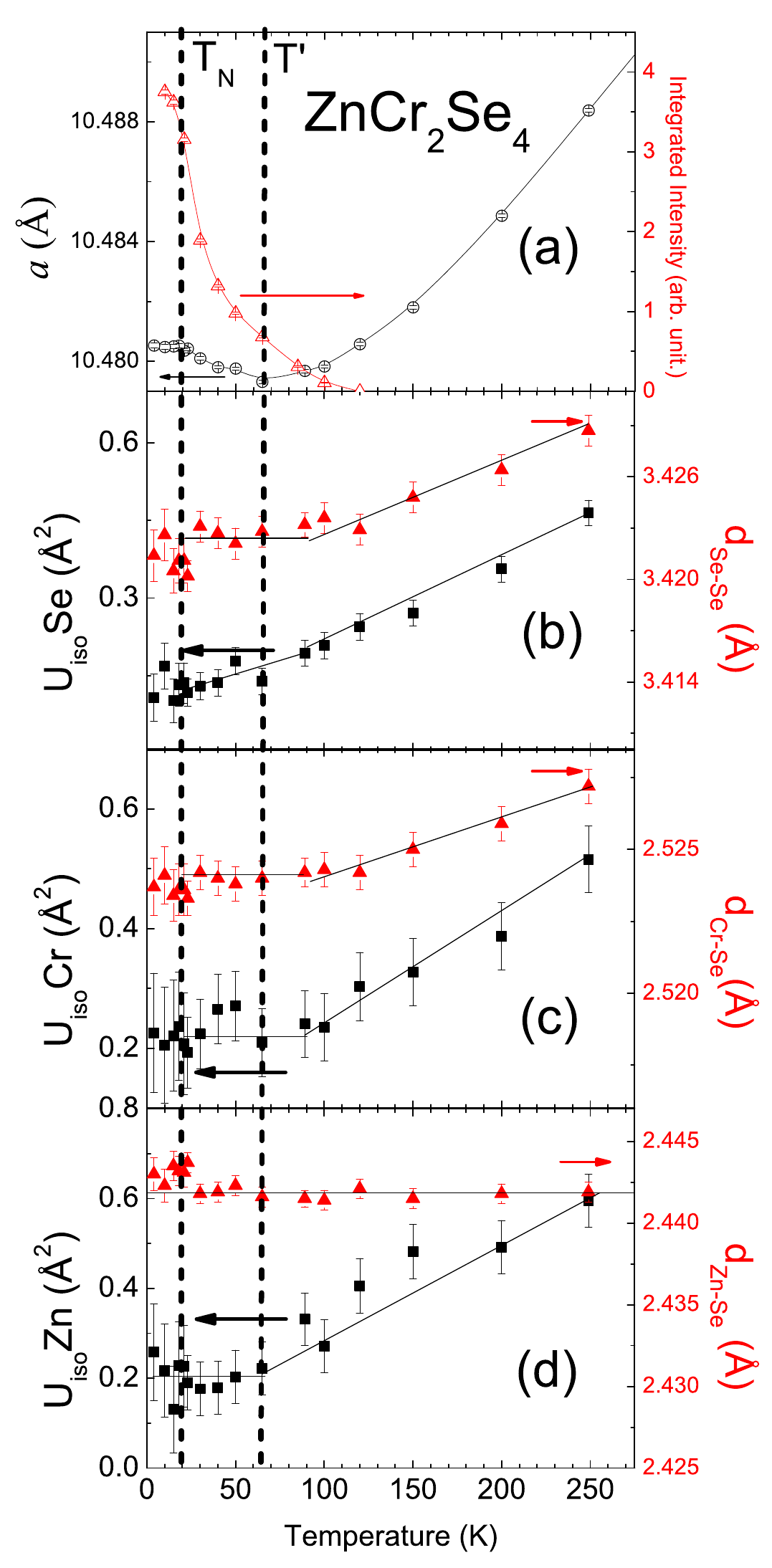}
\caption{\label{fig:fig2} (Color online) 
(a) Temperature dependence of the lattice constant \textit{a}
of the integrated diffuse magnetic intensity around the first magnetic 
reflection. The Se-Se, Cr-Se and Zn-Se bond lengths of the function of temperature
are shown in panels (b-c) respectively, together with Uiso values of
Se, Cr and Zn in the same panel respectively.
The lines
are drawn to guide the eye.\textbf{}}
\end{center}
\end{figure}

We now turn our attention to the structural parameters determined
by Rietveld analysis of the NPD data above $T_N$. We find that the
cubic lattice constant $a$ shows a positive thermal expansion
between 79~K and 300~K (Fig.~\ref{fig:fig2}(a)). This normal
behavior of the crystal lattice is also reflected in the Cr-Se and
Cr-Cr bond lengths, as well as in the isotropic atomic
displacement parameters ($U_{iso}$) of Cr and of Zn atoms (see
Fig.~\ref{fig:fig2}(b)-(d)), all of which decrease linearly over
this region in temperature. For the Zn-Se bond length we find an
unusual behavior in that it remains essentially temperature
invariant over the entire temperature range examined, showing a
slight increase below $T_N$. The latter feature may be indicative
of an orthorhombic distortion below $T_N$, as reported
elsewhere\cite{Hidaka03}. On further cooling below \Tp$\sim$~68~K
the cubic lattice constant exhibits NTE until $T_N$. Below \Tp\ we
also find that the $U_{iso}$ value of the Cr-atom, as well as the
Cr-Se bond length become temperature independent, further
suggesting an anomalous structural behavior. We did not detect a
structural phase transition at \Tp\, consistent with previous
structural investigations that report cubic symmetry for $T >
T_N$. This suggests that the observed structural behavior over the
temperature region $T_N < T < $~\Tp\ may originate from local
symmetry breaking only. Below $T_N$ and within the resolution of
our instrument we find an essentially constant value of the
lattice constant $a$. However, we believe that the refined
structural parameters of the NPD data reflect the average of a
slightly orthorhombically distorted structure, while our
measurements are insensitive to this small orthorhombic
distortion.

\begin{figure}[tb!]
\begin{center}
\includegraphics[scale=0.4]{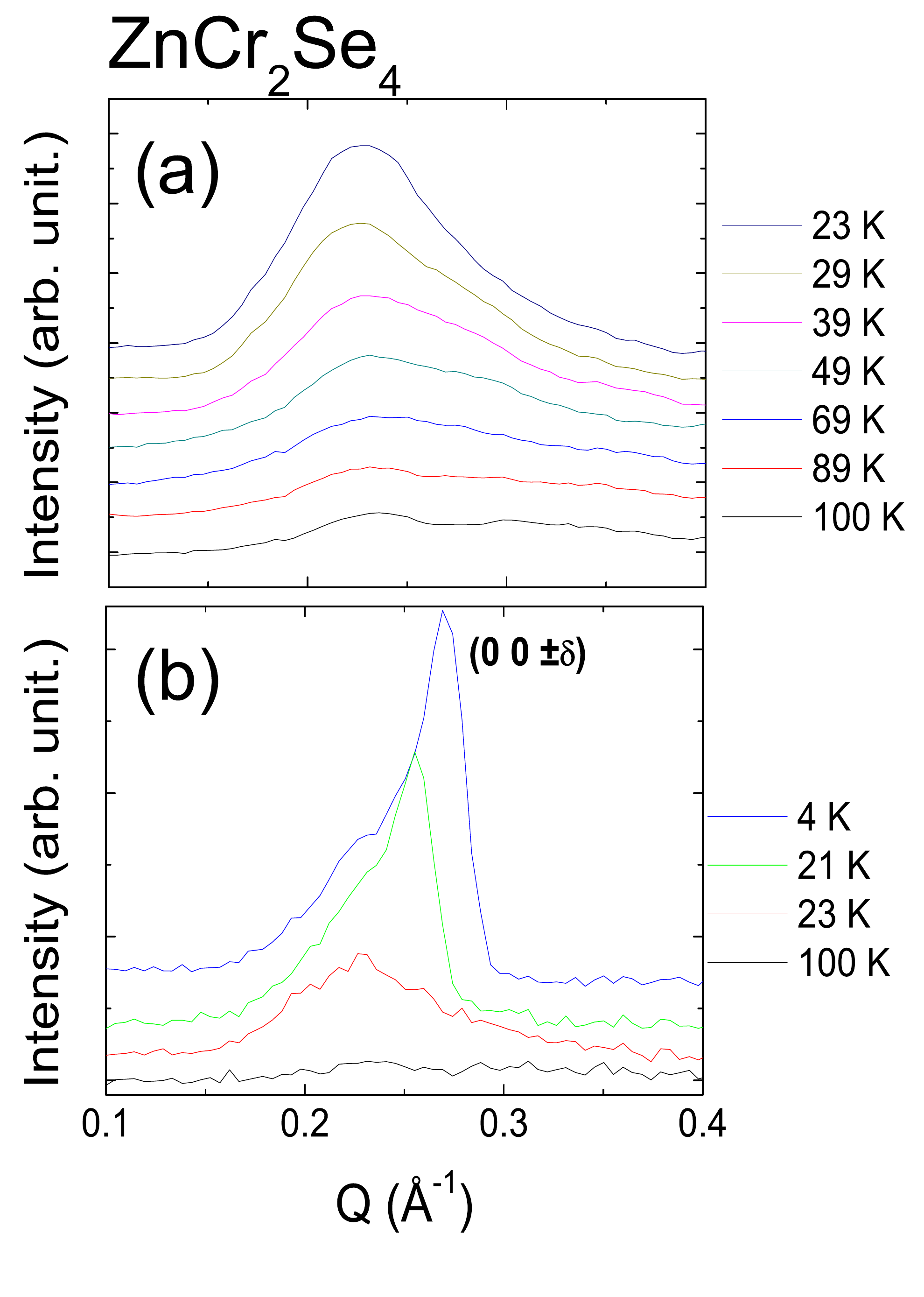}
\caption{\label{fig:meta1} (Color online) (a) Temperature dependence of the magnetic diffuse scattering
from 100 K to 23 K  from ZnCr$_2$Se$_4$. (b) The same region of Q-space showing the diffuse
magnetic scattering at 100~K and 23~K developing into the magnetic (0 0 $\pm\delta$) reflection
at 21 and 4 K below $T_N$.}
\end{center}
\end{figure}

The onset of the NTE may be closely related to the behavior of the
Cr-spins.  Specific heat and magnetization measurements
\cite{Hemberger07} suggest strong spin fluctuations below \Tp.
Indeed, broad diffuse scattering contributions centered around the
$(0,0,\delta)$ magnetic reflections are observed for temperatures
as high as 100~K. As illustrated in Fig.~\ref{fig:meta1}, this
diffuse scattering becomes narrower on cooling. Below $T_N=21$~K,
a sharp peak appears that is related to the helical magnetic
structure. The broad peak corresponding to magnetic fluctuations
coexists with the magnetic Bragg reflection within the long range
ordered ground state. This behavior is reminiscent of the complex
magnetic ordering process of the A-site spinel MnSc$_2$S$_4$
\cite{Krimmel06a} that was interpreted in terms of a spiral spin
liquid \cite{Bergman07}. The asymmetry of the broad peak is due to
the FM correlations within the $a-b$~planes. In
Fig.~\ref{fig:fig2}(a), we plot the integrated intensity of the
magnetic diffuse scattering around the $(0,0,\delta)$ reflection
as a function of temperature. This intensity appears above \Tp\
and rapidly increases as $T_N$ is approached. The simplest
plausible explanation of this scattering is that it arises from
small antiferromagnetic clusters with an average size of
$d\approx$ 53 \AA\ at $T\approx$ 23 K as given by the width (FWHM)
of the diffuse magnetic scattering.This observation is suggestive for a magneto-elastic coupling
above $T_N$ within the spin fluctuation regime, that leads to NTE
of the lattice below \Tp.

\subsection{High field NPD measurements of \zncrse}

\begin{figure}[tb!]
\begin{center}
\includegraphics[scale=0.6]{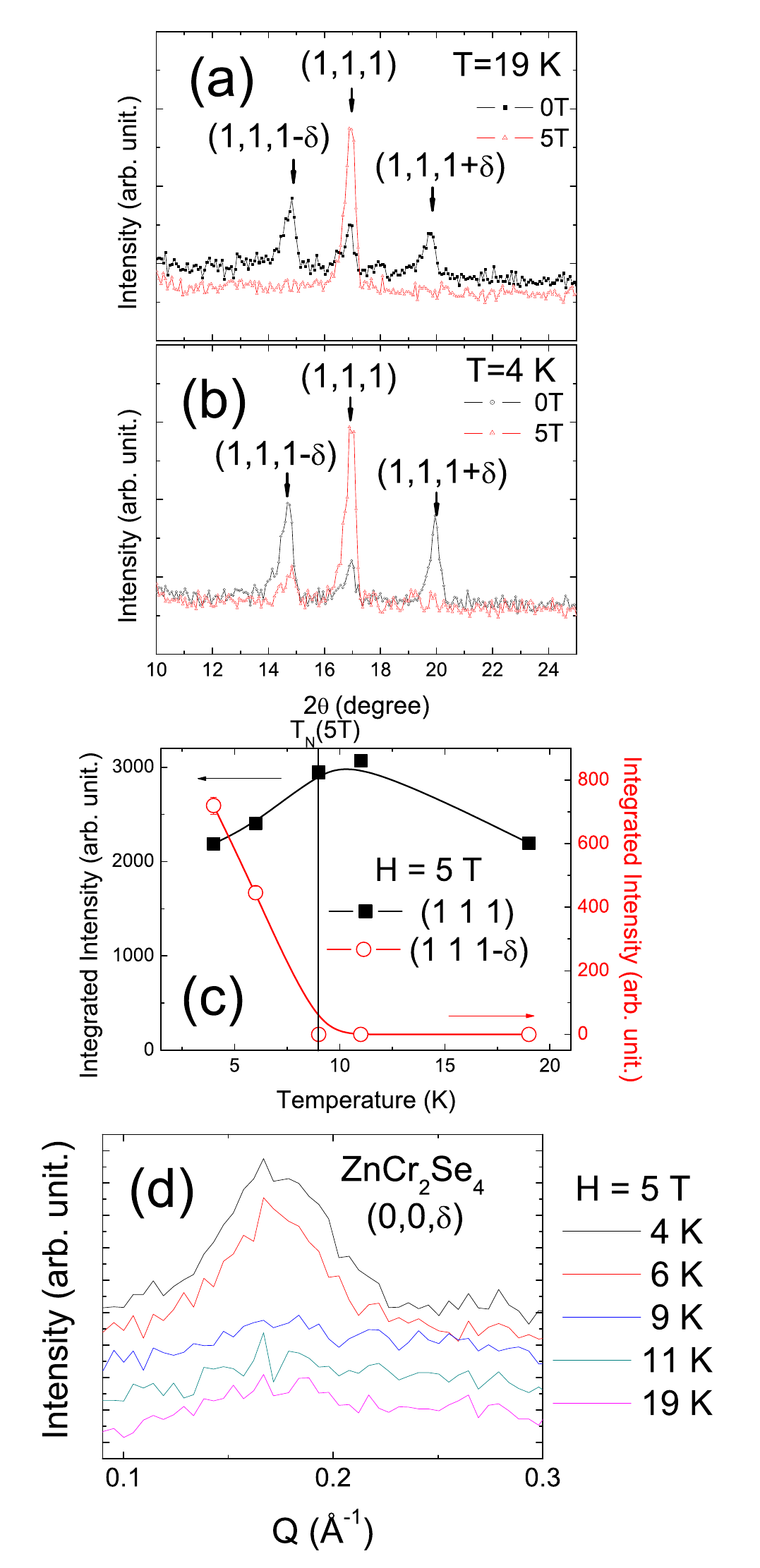}
\caption{\label{fig:fig6} (Color online) Low angle portion of the neutron
diffraction pattern of ZnCr$_2$Se$_4$ around the (1, 1, 1,)
reflection obtained in zero field (black line) and in an applied
magnetic field of 5 T (red) at (\textit{a}) 19 K and (\textit{b})
4 K. In panel (\textit{c}), the temperature dependence of the
integrated intensities of the nuclear Bragg reflection (1, 1, 1)
(black squares) and the corresponding magnetic satellite (1, 1,
1-$\delta$) (red circles) are shown. The vertical line indicates
the magnetic ordering temperature in a field of 5~T, where as the
dashed lines serve to guide the eye. Panel (d) shows the
magnetic (0 0 $\pm\delta$) Bragg reflection in a magnetic field of 5~T for
various temperatures. }\label{fig:fig6}
\end{center}
\end{figure}

We investigated the effect of a magnetic field on the crystal and
magnetic structure of \zncrse\ by measuring NPD data as a function
of temperature in a vertical magnetic field of 5T.  In this case,
magnetic scattering was observed only below 9~K indicating a
suppression of $T_N$ as compared to the zero field measurements.
Moreover, significant field induced changes of the magnetic
intensities are observed. These changes are summarized in
Fig.~\ref{fig:fig6}. At 19~K and 5~T we find a strong enhancement
of the nuclear (1,1,1) Bragg reflection compared to the zero field
data, while the helical (1,1,1$\pm\delta$) reflections are absent
(Fig.~\ref{fig:fig6}(a)). However, at the same temperature we find
broad diffuse intensity around the (0,0,$\delta$) reflection (see
Fig.~\ref{fig:fig6}(d)) indicative of magnetic short range helical
correlations (the correlation length is $\approx$ 42 \AA\ at 20
K). On further cooling, these short ranged helical correlations
developed into sharp magnetic Bragg reflections below 9~K (see
Fig.~\ref{fig:fig6}(b) and (c)) indicating that the 5~T ground
state is an induced ferromagnet with a weak residual AFM component
(with a correlation length at 4 K is $\approx$ 115 \AA ).

We now turn to the crystal structure of ZnCr$§_2$Se$_4$ in a
magnetic field of 5 T. The most interesting aspect is the
persistence of a NTE from 20~K (the highest temperature for an
applied magnetic field of 5~T) to 2~K.  In Fig.~\ref{fig:fig7} we
compare the temperature dependence of the cubic $a-$axis for both,
zero field and 5~T measurements in terms of the normalized lattice
constant $(a(T)-a(2K))/a(2K)$. The increase of the (average)
lattice constant appears to track the intensity of the AFM helical
Bragg reflections, similar as observed for zero field data (see
Fig.~\ref{fig:fig2} (a)).

\begin{figure}[tb!]
\begin{center}
\includegraphics[scale=0.3]{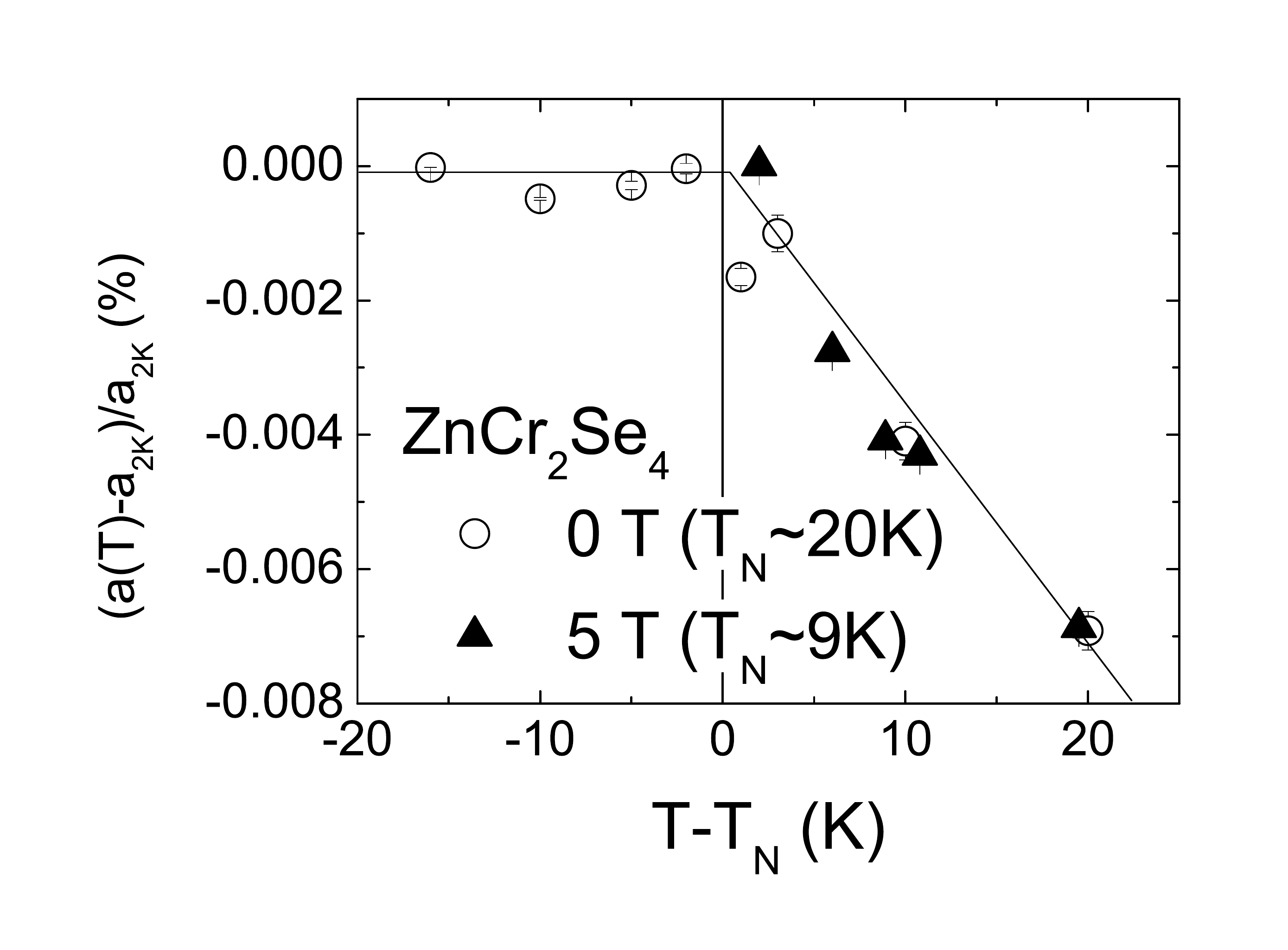}
\caption{\label{fig:fig7} Normalized lattice parameter \textit{a}
of ZnCr$_2$Se$_4$ versus reduced temperature T-$T_N$ at  0 T and  5 T,
respectively. Lines are drown to guide the eye.}\label{fig:fig7}
\end{center}
\end{figure}

\section{\label{ZnCrS} Crystal and magnetic structure of Z\lowercase{n}C\lowercase{r}$_2$S$_4$ }

\subsection{Synchrotron X-ray diffraction}

\begin{figure}[tb!]
\begin{center}
\includegraphics[scale=0.3]{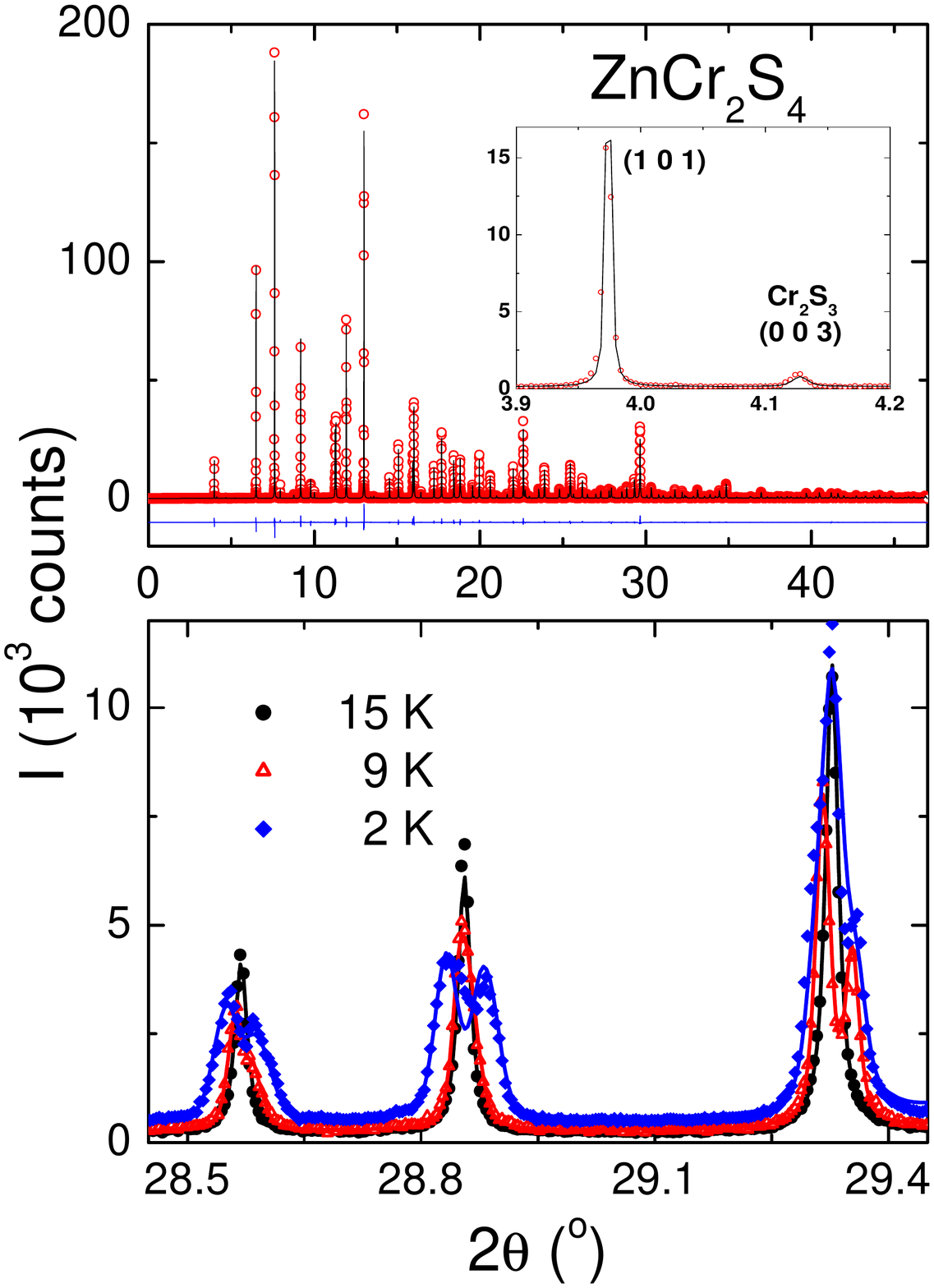}
\caption{(Color online) X-ray powder diffraction pattern of
ZnCr$_2$S$_4$. In the upper part, observed (red circles) and
calculated (black line) intensities and their difference (blue line
at the bottom) are shown for $T=9$~K. The inset shows the first
reflections (101) of ZnCr$_2$S$_4$ and (003) of Cr$_2$S$_3$ around
$2\theta \approx 4^o$ to illustrate the weak contribution of the
spurious phase with a respective volume fraction of about 2$\%$. The
lower frame shows the diffraction pattern of ZnCr$_2$S$_4$ for
various temperatures $T=2, 9$, and 15 K in a small angular range to
evidence the corresponding peak splitting across the two structural
phase transitions as described in the text. Again, symbols refer to
observed and lines to calculated profiles,
respectively.}\label{fig:fig4}
\end{center}
\end{figure}

We now turn to the crystal structure of \zncrs. In the upper panel
of Fig.~\ref{fig:fig4} we show an example of the XPD pattern
collected from our ZnCr$_2$S$_4$ sample. The sample was of
excellent quality as sharp, resolution limited diffraction peaks
could be observed up to extremely high $Q$-values. Due to the
extraordinarily high resolution, Bragg reflections appear as
vertical lines on this scale. The inset shows a strongly expanded
view around $2\theta \approx 4^o$ corresponding to the first
reflection of the diffraction pattern. Additionally to the
dominating (101) reflection of ZnCr$_2$S$_4$, now a weak intensity
of residual Cr$_2$S$_3$ appears. A magnification around 29$^o$
scattering angle evidences the different peak splitting due to the
structural phase transitions which are discussed below.

ZnCr$_2$S$_4$ passes through a first magnetic transition at
$T_{N1}=15$~K \cite{Hemberger06,Hamedoun86} which is accompanied
by a structural phase transition. For $T>15$~K in the paramagnetic
regime, ZnCr$_2$S$_4$ crystallizes in the normal cubic spinel
structure described by space group $Fd$\={3}$m$ in which the atoms
Zn, Cr and S are located at the Wyckoff positions 8$a$ , 16$d$ and 32$e$,
respectively. At 14 K, the crystal structure changes to tetragonal
symmetry of space group $I 4_1$/$amd$ (No. 141, origin choice 2) with half the unit cell
volume $V_{tetragonal}=a/\sqrt{2} \times a/\sqrt{2} \times a$ of
the cubic cell. The tetragonal crystal structure is observed in
the temperature range between the first and the second magnetic
transition $T_{N1} > T > T_{N2}$ (7 K - 14 K). In this structure,
the atoms Zn, Cr and S are found at the Wyckoff positions 4$a$ (0,
3/4, 1/8), 8$d$ (0, 0, 1/2) and 16$h$ (0, y, z), respectively.
Below 7 K, a further structural phase transition into orthorhombic
symmetry described by space group $Imma$ takes place. Apart from
the metric of the unit cell, the only structural parameters are
the sulfur positions since the Zn and Cr ions are always located
at special high-symmetry sites. For a given crystal structure, the
sulfur positional parameters remain virtually constant without any
significant temperature dependence. The same holds true for the
isotropic temperature factors of all three ion types that remain
almost constant with some scattering around $B_{iso} \approx
0.2$~(\AA$^2$). Only in the vicinity of $T_{N2}$ (specifically for
$T=5$~K, 7 K and 8~K) all temperature factors show enhanced values
of about 0.4 \AA$^2$.

\begin{table}[tb!]
\caption{\label{tab:table1}Crystallographic structure of
ZnCr$_2$S$_4$ as obtained from Rietveld refinements of high
resolution x-ray powder diffraction for $T=2$~K, 9 K and 17 K,
respectively. Listed are the lattice constants $a, b$ and $c$, the
sulfur positional parameters in fractional coordinates, the
isotropic temperature factors and the residuals of the refinement
$R_{Bragg}$ and $R_F$. Only the low temperature $Imma$ phase has
two different sulfur sites.}

\vspace{1ex}
\par

\begin{ruledtabular}
\begin{tabular}{ccccc}
ZnCr$_2$S$_4$ & $Imma$ & $I_1$/$amd$ & $Fd$\={3}$m$ \\
$T$(K) & 2 & 9 & 17\\
\hline

$a$~(\AA) &7.04679(5)&7.06116(2)&9.98152(2)\\
$b$~(\AA) &7.06743(5)&&\\
$c$~(\AA) &9.98328(8)&9.97137(4)&\\
$x$(S1) &0&0&0.25972(3)\\
$x$(S2) &0.26136(9)&&\\
$y$(S1) &0.52253(18)&0.019567(13)&0.25970(3)\\
$y$(S2) &0.25&&\\
$z$(S1) &0.26136(9)&0.25955(11)&0.25970(3)\\
$z$(S1) &-0.01136(9)&&\\
$B_{Zn}$ (\AA$^2$) &0.131(13)&0.203(8)&0.127(6)\\
$B_{Cr}$ (\AA$^2$) &0.189(39)&0.200(9)&0.126(6)\\
$B_{S}$ (\AA$^2$) &0.184(17)&0.221(9)&0.135(7)\\
$R_{Bragg} (\%)$ &4.36&3.48&3.69\\
$R_F (\%)$ &4.72&3.52&3.89\\
\hline

\end{tabular}
\end{ruledtabular}
\end{table}

\begin{figure}[tb!]
\begin{center}
\includegraphics[scale=0.3]{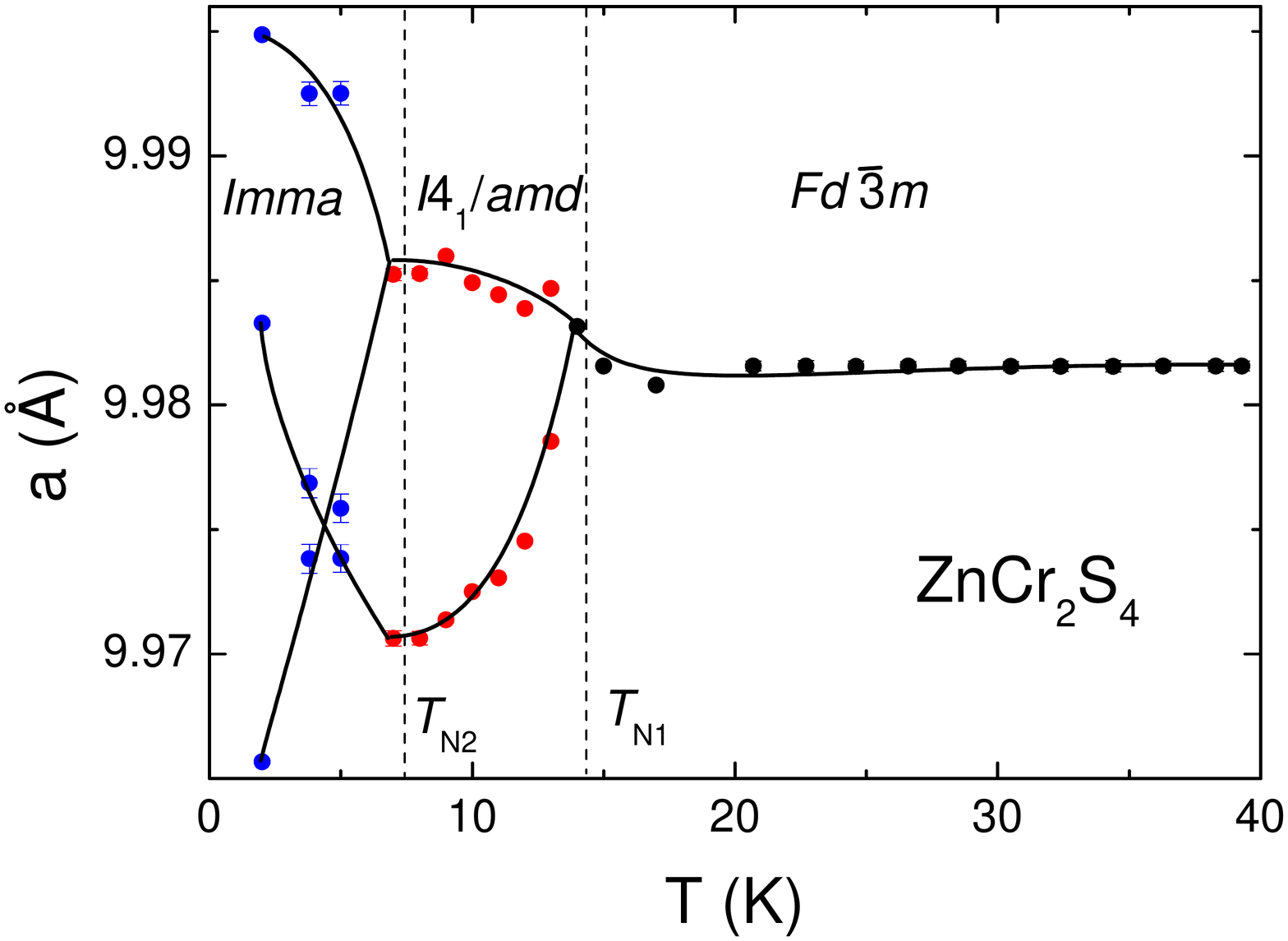}
\caption{(Color online) Temperature dependent normalized lattice
constants of ZnCr$_2$S$_4$. The magnetic transition temperatures
and the space groups of the different structures are indicated.
The solid lines are to guide the eye. }\label{fig:fig5}
\end{center}
\end{figure}

The results of the structure refinements are detailed in Table 1
for selected temperatures $T=2$~K at base temperature
(orthorhombic phase), $T=9$~K (within the tetragonal phase) and 17
K (cubic phase). Fig.~\ref{fig:fig5} shows the temperature
dependence of the lattice constants of ZnCr$_2$S$_4$. For a direct
comparison, the lattice parameters $a$ (and $b$) of the tetragonal
(orthorhombic) phases have been normalized by multiplication with
$\sqrt{2}$. Evidently, the magnetic transitions are directly
followed by changes of the crystal structure.

\subsection{Zero field NPD measurements of \zncrs}

\begin{figure}[tb!]
\begin{center}
\includegraphics[scale=0.3]{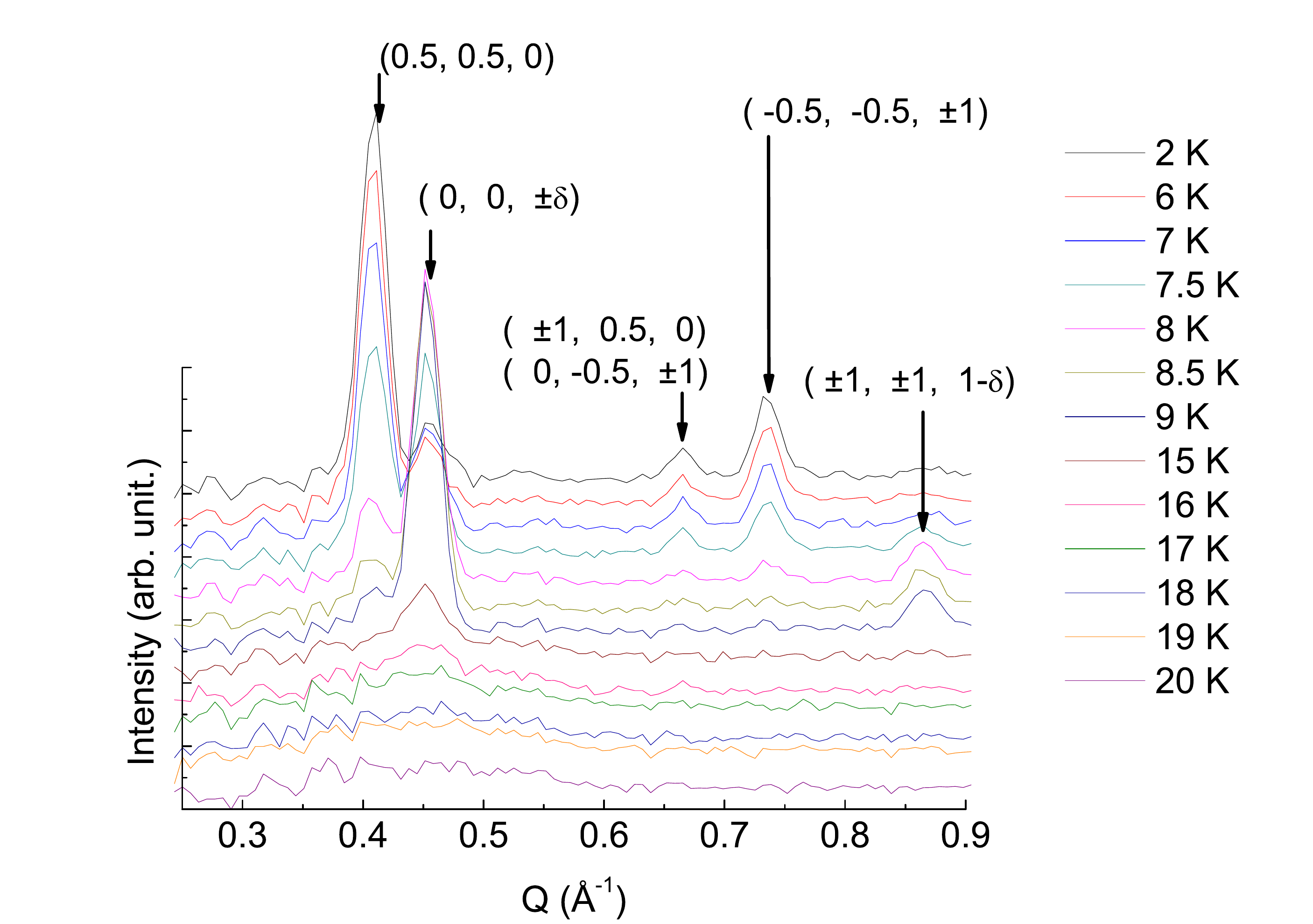}
\caption{ (Color online) Temperature dependence of the first magnetic peaks of
ZnCr$_2$S$_4$ with propagation vectors $k_1=(0, 0, 0.787)$, $k_2=(1, 0.5,0)$ and $k_3=(0.5, 0.5, 0)$).
Measurements were performed on E6 with
$\lambda$=2.446$\AA$.}\label{fig:metaZnCrS}
\end{center}
\end{figure}

The two antiferromagnetic transitions of ZnCr$_2$S$_4$ at
$T_{N1}=15$~K and $T_{N2}=8$~K are also reflected in the
temperature dependent NPD data shown in Fig.~\ref{fig:metaZnCrS}.
Below $T_{N1}$ at 14K we observe 12 magnetic peaks at low-$Q$, all
of which can be indexed by the propagaton vector k$_1\sim(0, 0,
\delta)$ with $\delta=$0.787(1). This ordering is very similar to
what we found for the \zncrse\ compound
\cite{Plumier75,Hamedoun86}. Here, our NPD data can be modeled using the
same FM ordering in (001) planes with a screw angle between planes
of 70.8(1)\dg at 14K and a Cr magnetic moment of 1.23(3) $\mu_B$,
similar to the results of reference \onlinecite{Hamedoun86}.

\begin{figure}[tb!]
\begin{center}
\includegraphics[scale=0.3]{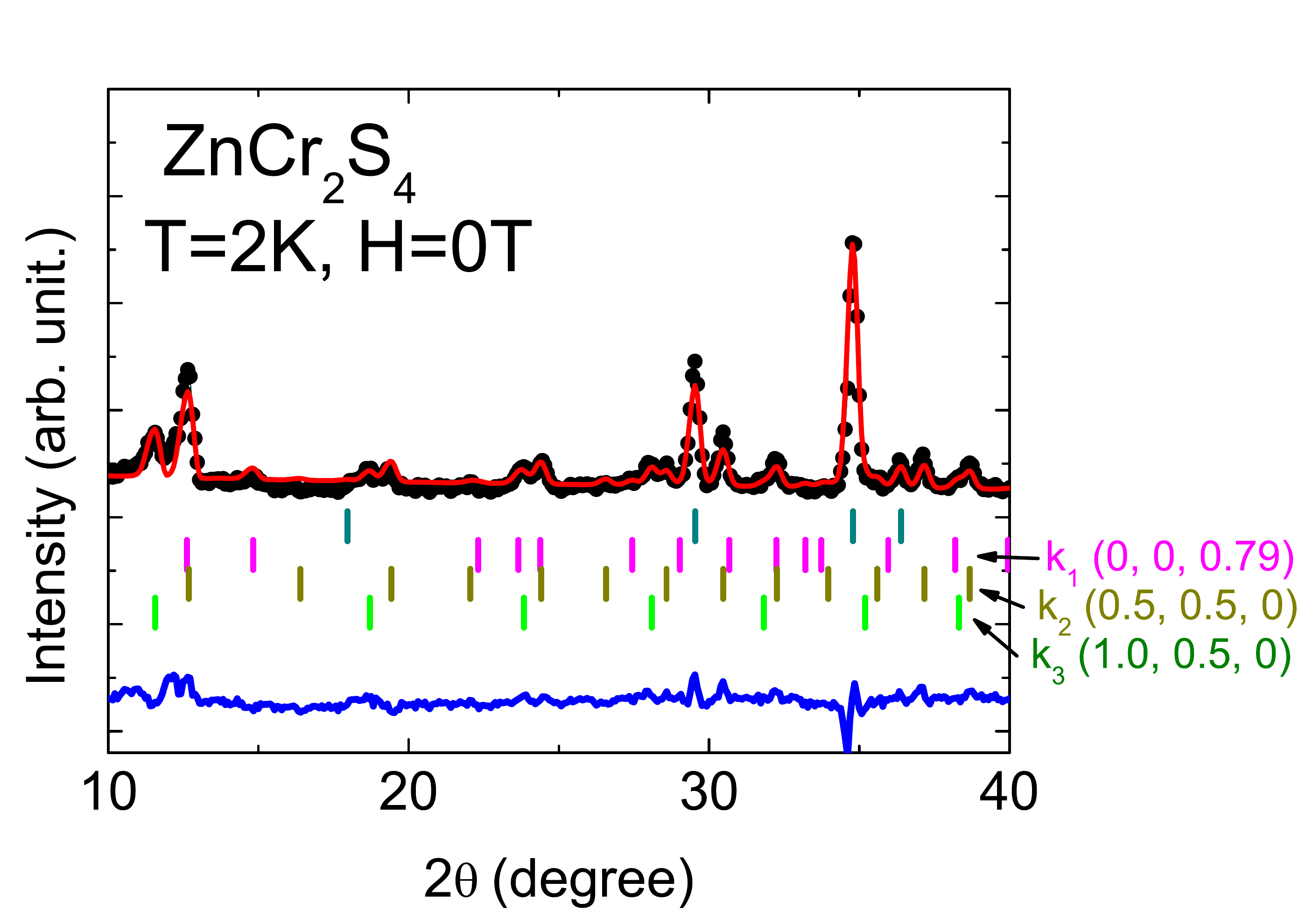}
\caption{ (Color online) Rietveld analysis of NPD data measured from
 ZnCr$_2$S$_4$. The black points represent
the experimental data while the red and blue lines represent the
model and the difference between the experimental data and
the model respectively. The tick marks show the expected positions of nuclear
Bragg reflections and magnetic reflections from the three magnetic orderings 
described by propagation vectors $k_1$, $k_2$ and $k_3$.Here we have considered that the three propagation vectors are in three spatially separated magnetic phases. }\label{fig:fig3}
\end{center}
\end{figure}

Below \Tnt\ the ordering of Cr-spins is more complex. On cooling,
the intensity of the $k_{1}$ reflections decreases and below 8K
new commensurate reflections appear (Fig.~\ref{fig:metaZnCrS})
with propagation vectors  $k_2=(0.5, 0.5, 0)$ and $k_3=(1.0, 0.5, 0)$. Indeed the same behavior in the magnetic structure of ZrCr$_{2}$S$_{4}$ was noted in previous work\cite{Hamedoun86}. 

Modeling the magnetic structure of ZrCr$_{2}$S$_{4}$ below \Tnt\ is especially challenging considering that the symmetry of the $k_{2}$ and $k_{3}$ magnetic wave vectors in general is not exclusively orthogonal to $k_{1}$. This is certainly the case for $k_{2}$, however $k_{3}$ can describe a collinear magnetic structure with ferromagnetic planes perpendicular to $k_{3}$ which is arranged as $+ + - -$ sequence along the propagation vector and orthogonal to $k_{1}$. Hamedoun \etal\ argued that such apparently compicated magnetic structure reflects the chemical complexity of the samples\cite{Hamedoun86}. The variation of the magnetic components for the three different
wave vectors in different samples prepared under different partial
pressure of S, lead to the conclusion that the apparent complexity
in the magnetic ordering arises from S-vacancies
\cite{Hamedoun86}, the differences between powder and single
crystal samples being especially pronounced. Such behavior is
proposed to arise due to the sensitivity of the exchange
parameters to the Cr-Cr distance which is in turn are modulated by
S-vacancies. A calculation of the free energy between the possible
magnetic structures represented by $k_{1}$ and $k_{2}$ differs by
only $\sim$0.1K \cite{Hamedoun86} further highlighting the virtual
degeneracy of these states. Application of the same magnetic model
used by Hamedoun \etal\ to our NPD data measured at 2K yielded
similar results. Here we deduce $m_{1}=$1.13(2), $m_2$=1.18(2) and
$m_3$=0.69(4)~$\mu_B/Cr$, compared to 1.23, 1.53 and 0.85
$\mu_B/Cr$ respectively \cite{Hamedoun86}. 

If indeed the magnitude of these magnetic moments is dependent on S inhomogeneities it is peculiar that our results would be so similar to those of reported previously. To test further possible solutions to the magnetic structure of ZrCr$_{2}$S$_{4}$ below \Tnt\ assuming a single magnetic phase, we preformed symmetry analysis and phenomenological fits to the NPD data on the basis of the low temperature orthorhombic phase. Our working assumption was to test if the magnetic scattering described by $k_{2}$ and $k_{3}$ can arise from components of the Cr moment that are perpendicular to the ferromagnetic planes described by the $k_{1}$ wave vector. 

For $k_{2}$, symmetry analysis presents the possibility of components of the Cr moment that are parallel to the $c-$axis (i.e. orthogonal to Cr moments described by $k_{1}$). Poor fits to the NPD data were obtained  for  $k_{2}$ Bragg peaks when constraining the Cr moment to be parallel to the $c$-axis. Better results, however were obtained when the Cr moments were allowed to have components in the [110] plane, clearly demonstrating that $k_{1}$ and $k_{2}$ spins are not orthogonal and therefore can not arise from the same chemical phase. A similar analysis for the $k_{3}$ propagation vector shows that the magnetic intensity of these Bragg reflections can be modeled only with Cr spin components that are parallel to the $c-$axis, indicating a compatibility between the $k_{1}$ and $k_{3}$ wave vectors. Assuming such a configuration and analyzing the NPD data with two magnetic phases ($k_{1}+k_{3}$ and $k_{2}$) we  obtained $m_{1+3}=$1.63(2) and $m_2$=2.16(2). However, we note that a thorough search of the NPD data could not identify Bragg reflection with $k=k_{1}+k_{2}=(1, 0.5, \delta)$, suggesting again that $k_{1}$ and $k_{3}$ likely do not arise from the same chemical phase. 

To make further progress in describing the true magnetic ground state of ZrCr$_{2}$S$_{4}$ single crystal measurements from stoichiometric samples are required, an endeavor  beyond the scope of the present work.

\subsection{High field NPD measurements of \zncrs}

\begin{figure}[tb!]
\begin{center}
\includegraphics[scale=0.3]{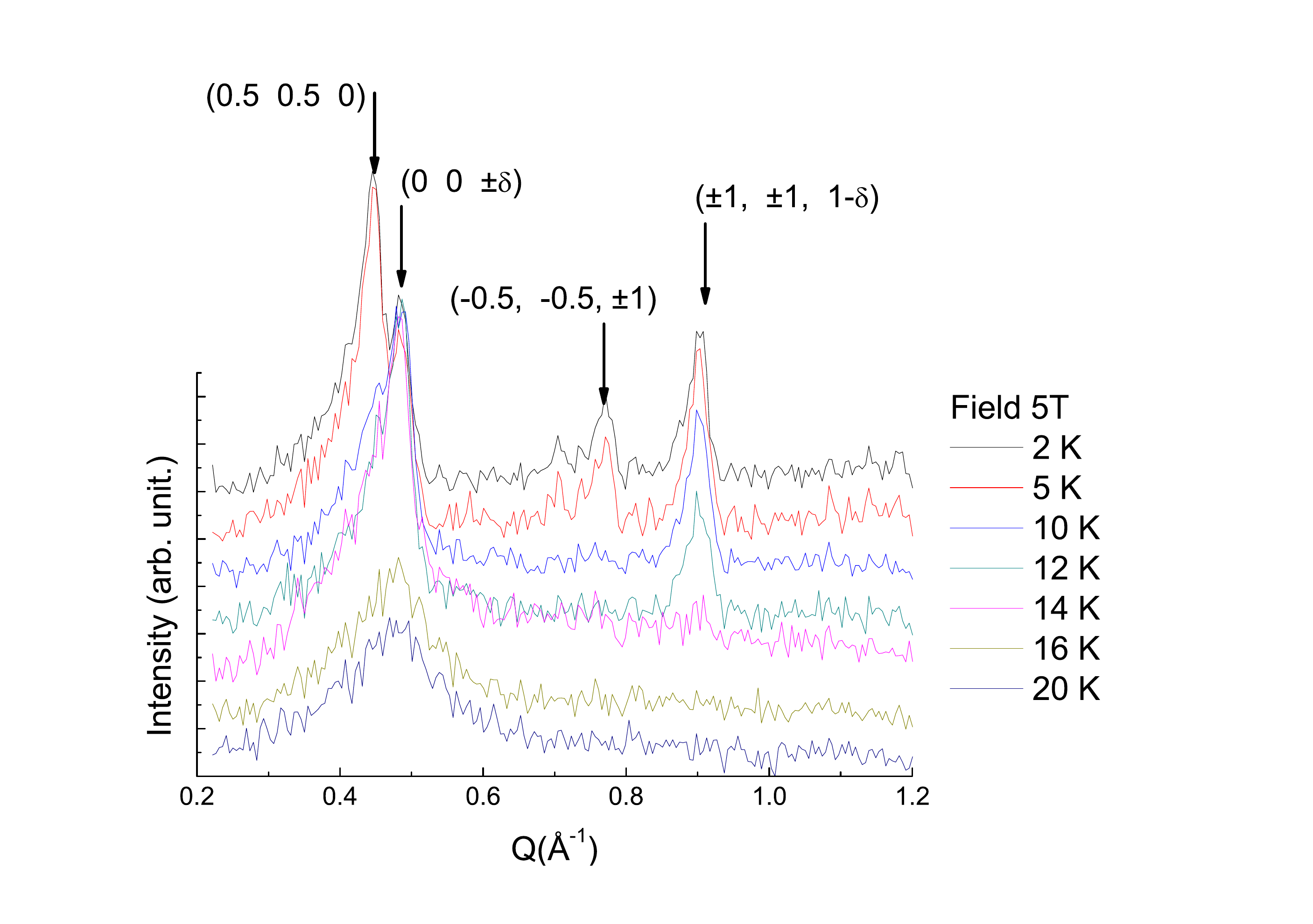}
\caption{ (Color online) Low angle magnetic reflections of ZnCr$_2$S$_4$ in an
applied magnetic field of 5~T for various temperatures. NPD data were
measured on E9 with $\lambda$=1.797$\AA$.}\label{fig:meta5TS}
\end{center}
\end{figure}

Given the degeneracy between the different magnetic orderings in
\zncrs, the application of a magnetic field offers the possibility
of energetically favoring one of these states. Indeed, NPD as a
function of temperature in an applied field of 5T indicates that
the helical magnetic structure ($k_1$) is preferred at the expense
of the two commensurate orderings ($k_{2}$ and $k_{3}$). This
behavior is evidenced in Fig.~\ref{fig:meta5TS}. At 20~K and 5~T,
the magnetic scattering appears as a broad diffuse reflection
indicative of AFM fluctuations. On cooling, this scattering
becomes sharper indicative of a long range helical magnetic
ordering below $T_N(5~T)=14$~K. The magnetic structure at 10 K
described by the incommensurate wavevector $k_{1}$=(0,0,0.78(1))
gives a helical angle for the Cr-spins of 70.2(9)\dg\ . On
further cooling below 5 K, the commensurate reflections with
propagation vectors $k_{2}$ and $k_{3}$ appearwhile the intensity of
the commensurate reflections is significantly suppressed compared
to the zero field data. Thus a 5T field leads to a ground state that is similar
to that of \zncrse\ at zero field albeit with additional small
commensurate components, $k_{2}$ and $k_{3}$.

\begin{figure}[tb!]
\begin{center}
\includegraphics[scale=0.3]{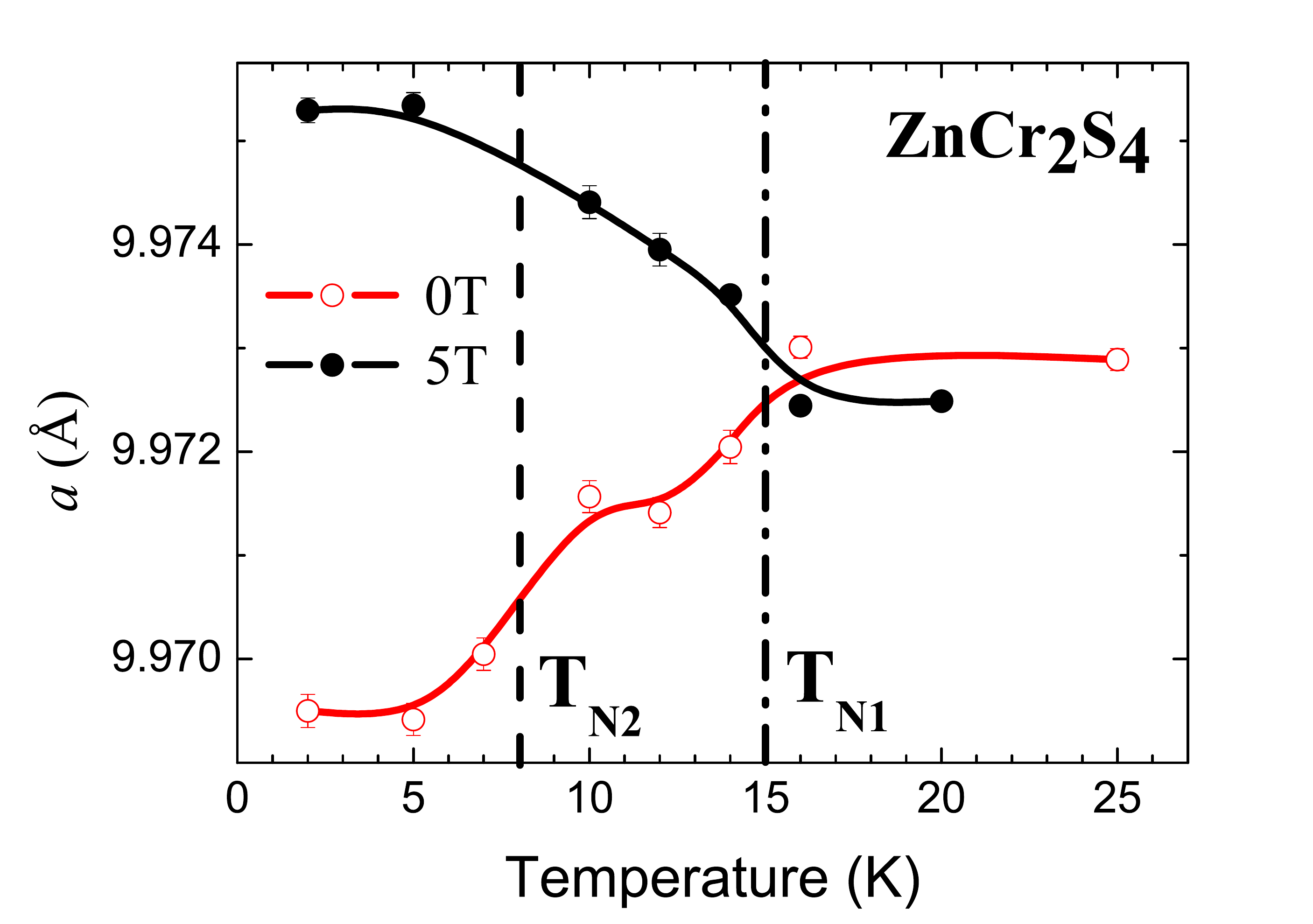}
\caption{ (Color online) Comparison of the temperature dependence
of the lattice parameter \textit{a} for a magnetic field of
5~T (black line and full circles) and in zero field (red line and
open circles). Lines are drawn to guide the
eye.}\label{fig:fig9}
\end{center}
\end{figure}

The dominant helical spin ordering at 5~T is also evidenced in the
thermal expansion. In Fig.~\ref{fig:fig9} the temperature
dependence of the average cubic lattice constant in zero field and
5~T is shown. An overall positive thermal expansion in zero field
is observed while the 5~T data reveal a NTE
coinciding with the onset of helical ordering.  This behavior is
reminiscent of the zero field measurements of the \zncrse\
samples. It may suggest that the helical spin ordering induces
frustration onto the lattice that leads to anomalous thermal
expansion, as well as anomalies in the phonon spectra
\cite{Hemberger07}.

\section{\label{sec:disc}Discussion}

Spin-lattice coupling has turned out to be an efficient way to
lift frustration. Therefore, the magnetic structure and structural
distortions in chromium spinels are strongly correlated.
ZnCr$_2$O$_4$ exhibits a small Cr-Cr distance and, consequently, a
strong direct AFM exchange as reflected in a large negative
Curie-Weiss temperature. Below $T_N=12.5$~K it orders
antiferromagnetically in a commensurate, but complex multi-{\bf k}
structure whose details depend on sample preparation conditions
\cite{Lee07}. The accompanying structural phase transition is
characterized by a contraction of the $c$-axis and structural
superlattice peaks of type (1/2, 1/2, 1/2) corresponding to the
tetragonal space group $I$\={4}$m2$ \cite{Lee07}. CdCr$_2$O$_4$
has somewhat larger Cr-Cr distance but still dominating AFM
exchange. It reveals an incommensurate spin structure with
propagation vector ${\bf k}=(0, \approx 0.09, 1)$ and an elongated
$c$-axis within tetragonal symmetry of space group $I4_1$/$amd$.
On the other hand, ZnCr$_2$Se$_4$ with the largest Cr-Cr
separation has strong FM exchange evidenced by a large positive
Curie-Weiss temperature of $\theta_{CW}=115$~K. A structural phase
transition takes place at $T_N=21$~K from cubic $Fd$\={3}$m$ to
orthorhombic $Fddd$ symmetry with a contraction of the $c$-axis
\cite{Hidaka03}. ZnCr$_2$Se$_4$ reveals an incommensurate helical
spin structure with FM layers and AFM coupling along the
orthorhombic $c$ axis of rotation. For ZnCr$_2$S$_4$, both FM and
AFM interactions are almost equal in strength leading to a
Curie-Weiss temperature close to zero \cite{Rudolf07}. The subtle
balance between two competing interactions is easily disturbed and
therefore a strong sensitivity of the magnetic properties on
slight structural distortions is expected. Neutron diffraction
revealed the onset of a helical spin structure at $T_{N1}$. Below
12 K, the helical structure transforms into two collinear
structures with wave vectors [1/2, 1/2, 0] and [1, 1/2, 0]
\cite{Hamedoun86}. At low temperatures, a coexistence of all three
phases was suggested. Along the magnetic structure determination,
an upper limit of $1-c/a<0.002$ has been reported for any static
structural distortions \cite{Hamedoun86}. The magnetic structure
of ZnCr$_2$S$_4$ below $T_{N1}=15$~K is similar to that of the Se
homologue, formed by dominating FM interactions leading to FM
layers. To account for the additional AFM exchange, a helix is
formed with a rotation axis perpendicular to [110]. Upon further
cooling, the relative weight of the AFM exchange increases and the
helix transforms into commensurate AFM structures reminiscent of
ZnCr$_2$O$_4$ with identical propagation vectors. A phase
coexistence has been proposed based on the fact that significant
differences in the diffraction intensities between powder and
single crystalline samples were observed \cite{Hamedoun86}. In
view of the sensitivity of the magnetic properties on sample
preparation conditions (e. g. vacancies) such a conclusion has to
be considered with care. Our present high resolution x-ray
diffraction study revealed structural phase transitions from cubic
$Fd$\={3}$m$ to tetragonal $I4_1$/$amd$ symmetry just below
$T_{N1}$ and from tetragonal $I4_1$/$amd$ to orthorhombic $Imma$,
just below $T_{N2}$, but no anomalous behavior around 12 K. For
each temperature region, a single phase of ZnCr$_2$S$_4$ was
sufficient to account for the measured diffraction pattern. It is
interesting to note that the orthorhombic low temperature
structure of ZnCr$_2$S$_4$ is identical to that of magnetite below
the Verwey transition with mutually perpendicular rows of ferric
and ferrous ions along the [110] directions \cite{Hamilton58}. A
corresponding scenario in case of ZnCr$_2$S$_4$ would be that
below $T_{N2}=8$~K, the helix of FM spin chains adopts a phase of
$\pi/2$ leading to a spin arrangement of perpendicular FM spin
chains along [110]. This would correspond to a magnetic analogue
of the charge order process in Fe$_3$O$_4$. Regardless other measurements
on magnetic field in ZnCr$_2$S$_4$ clearly demonstrate the greater stability of the helical
ordering compared to the commensurate $k_{2}$ and $k_{3}$ orderings.

The formation of ferromagnetic layers in ZnCr$_2$Se$_4$ evidenced by the
magnetic fluctuations below 80 K, as shown in Fig.~\ref{fig:meta1} at 0 T,
induces frustration in the lattice parameter. One can argue that such behavior
may be related to the observation of a NTE.In an applied magnetic field, the helical spin structure of
ZnCr$_2$Se$_4$ becomes suppressed on increasing FM interactions,
moving down the magnetic transition temperature and the
temperature range of NTE. On the other hand, in case of
ZnCr$_2$S$_4$, the magnetic field starts to break the delicate
balance between AFM and FM interactions. An external magnetic
field amplifies the FM interactions and induces a NTE behavior.
ZnCr$_2$S$_4$ differs from the selenide compound as the helical
spin structure is not suppressed in an applied magnetic field.

\section{\label{sec:concl}Conclusion:}

In this paper we have presented a comparison of two closely
related spinel compounds ZnCr$_2$Se$_4$ and ZnCr$_2$S$_4$. They
reveal different magnetic structures originating from the subtle
interplay between frustrated FM and AFM exchange interactions. In
the selenide compound, we relate the evolution of incommensurate
helical magnetic correlations and NTE as a consequence of magnetic
frustration. The structural effects by an applied magnetic field
can be accounted for accordingly by enhancing the FM component
with respect to the AFM exchange. Therefore, the magnetic ordering
temperature and the range of NTE is shifted towards lower
temperatures in agreement with bulk measurements
\cite{Hemberger07} and a new magnetic structure (conical/cyclodal
structure) is induced, as predicted in the literature
\cite{Murakawa08}.

The thio-spinel ZnCr$_2$S$_4$ exhibits two magnetic transitions.
Similar to ZnCr$_2$Se$_4$ \cite{Hidaka03,Hemberger07}, the
magnetic transitions of ZnCr$_2$S$_4$ are accompanied by
structural transformations: from cubic ($Fd$\={3}$m$) to
tetragonal ($I 4_1$/$amd$) at $T_{N1}=15$~K, and from
tetragonal($I 4_1$/$amd$) to orthorhombic ($Imma$) at
$T_{N2}=8$~K. Applying an external magnetic field induces NTE
below the first magnetic transition temperature and further
suppresses the commensurate magnetic structural components in the
second low temperature magnetic phase of ZnCr$_2$S$_4$.

\acknowledgments

This work was supported by the Deutsche Forschungsgemeinschaft DFG
via SFB 484, Augsburg. The help of M. M\"ucksch during the
synchrotron diffraction experiments is acknowledged. We also acknowledge the support of
BENSC in providing the neutron research facilities used in this work.



\begin{thebibliography}{99}


\bibitem{Rudolf07} T. Rudolf, Ch. Kant, F. Mayr, J. Hemberger,
V. Tsurkan, and A. Loidl, New J. Phys. {\bf 9}, 76 (2007).

\bibitem{Canals98} B. Canals and C. Lacroix, Phys. Rev. Lett. {\bf
80}, 2933 (1998).

\bibitem{Ramirez99} A. P. Ramirez, A. Hayashi, R. J. Cava, R. Siddhartan, and
B. S. Shastry, Nature {\bf 399}, 333 (1999).

\bibitem{Bramwell01} S. P. Bramwell and M. J. P. Gringas, Science {\bf
249}, 1495 (2001).

\bibitem{Garcia00} A. J. Garcia-Adeva and D. L. Huber, Phys. Rev.
Lett. {\bf 85}, 4598 (2000).

\bibitem{Radaelli02} P. G. Radaelli, Y. Horibe, M. J. Gutmann, H.
Ishibashi, C. H. Chen, R. M. Ibberson, Y. Koyama, Y.-S. Hor, V.
Kiryukhin, and S.-W. Cheong, Nature {\bf 416}, 155 (2002).

\bibitem{Lee02} S.-H. Lee, C. Broholm, W. Ratcliff, G. Gasparovic,
Q. Huang, T. H. Kim, and S.-W. Cheong, Nature {\bf 418}, 856
(2002).

\bibitem{Kondo97} S. Kondo, D. C. Johnston, C. A. Swenson, F. Borsa, A. V. Mahajan,
L. L. Miller, T. Gu, A. I. Goldman, M. B. Maple, D. A. Gajewski,
E. J. Freeman, N. R. Dilley, R. P. Dickey, J. Merrin, K. Kojima,
G. M. Luke, Y. J. Uemura, O. Chmaissem, and J. D. Jorgensen, Phys.
Rev. Lett. {\bf 78}, 3729 (1997).

\bibitem{Krimmel99} A. Krimmel, A. Loidl, M. Klemm, S. Horn, and H.
Schober, Phys. Rev. Lett. {\bf 82}, 2919 (1999).

\bibitem{Berg03} E. Berg, E. Altman, and A. Auerbach, Phys. Rev.
Lett. {\bf 90}, 147204 (2003).

\bibitem{Reehuis03} M. Reehuis, A. Krimmel, N. B\"uttgen, A. Loidl, and A.
Prokofiev, Europ. Phys. J. B {\bf 35}, 311 (2003).

\bibitem{Tschernyshyov02} O. Tchernyshyov, R. Moessner, and L. Sondhi,
Phys. Rev. Lett. {\bf 88}, 067203 (2002).

\bibitem{Yamashita00} Y. Yamashita and K. Ueda, Phys. Rev. Lett. {\bf 85},
4960 (2000).

\bibitem{Lee00} S.-H. Lee, C. Broholm, T. H. Kim, W. Ratcliff II,
and S.-W. Cheong, Phys. Rev. Lett. {\bf 84}, 3718 (2000).

\bibitem{Sushkov05} A. B. Sushkov, O. Tchernyshyov, W. Ratcliff
II, S.-W. Cheong, and H. D. Drew, Phys. Rev. Lett. {\bf 94},
137202 (2005).

\bibitem{Chung05} J.-H. Chung, M. Matsuda, S.-H. Lee, K. Kakurai,
H. Ueda, T. J. Sato. H. Takagi, K.-P. Hong, and S. Park, Phys.
Rev. Lett. {\bf 95}, 247204 (2005).

\bibitem{Baltensperger68} W. Baltensperger and J. S. Helman, Helv.
Phys. Acta {\bf 41}, 668 (1968).

\bibitem{Baltensperger70} W. Baltensperger, J. Appl. Phys. {\bf
41}, 1052 (1970).

\bibitem{Bruesch72} P. Br\"uesch and F. D\'~Ambrogio, Phys. Stat.
Sol. b {\bf 50}, 513 (1972).

\bibitem{Massidda99} S. Massidda, M. Posternak, A. Baldereschi,
and R. Resta, Phys. Rev. Lett. {\bf 82}, 430 (1999).

\bibitem{Fennie06} C. J. Fennie, and K. M. Raabe, Phys. Rev. Lett.
{\bf 96}, 205505 (2006).

\bibitem{Hemberger06} J. Hemberger, T. Rudolf, H.-A. Krug von
Nidda, F. Mayr, A. Pimenov, V. Tsurkan, and A. Loidl, Phys. Rev.
Lett. {\bf 97}, 087204 (2006).

\bibitem{Plumier66} R. Plumier, J. Phys. (Paris) {\bf 27}, 213
(1966).

\bibitem{Hemberger07} J. Hemberger, H.-A. Krug von
Nidda, V. Tsurkan, and A. Loidl, Phys. Rev. Lett. {\bf 98}, 147203 (2007).

\bibitem{Kleinberger66} R. Kleinberger and R. de Kouchkovsky, CR Acad.
Sci. (France) {\bf 262}, 628 (1966).

\bibitem{Hidaka03} M. Hidaka, N. Tokiwa, M. Fuji, S. Watanabe, and
J. Akimitsu, Phys. Stat. Sol. (b) {\bf 236}, 9 (2003).

\bibitem{Rudolf207} T. Rudolf, Ch. Kant, F. Mayr, J. Hemberger,
V. Tsurkan, and A. Loidl, Phys. Rev. B {\bf 75}, 052410 (2007).

\bibitem{Kimura03} T. Kimura, T. Goto, H. Shintani, K. Ishizaka,
T. Arima, and Y. Tokura, Nature (London) {\bf 426}, 55 (2003).

\bibitem{IOP} For reviews see: M. Fiebig, J. Phys. D {\bf 38},
R123 (2005); S.-W. Cheong and M. Mostovoy, Nature Mat. {\bf 6}, 13
(2007); J. Phys.: Cond. Mat., in print (2008).

\bibitem{Yamasaki06} Y. Yamasaki, S. Miyasaka, Y. Kaneko, J. P.
He, T. Arima, and Y. Tokura, Phys. Rev. Lett. {\bf 96}, 207204
(2006).

\bibitem{Katsura05} H. Katsura, N. Nagaosa, and A. V. Balatsky,
Phys. Rev. Lett. {\bf 95}, 057205 (2005).

\bibitem{Sergienko06} I. A. Sergienko and E. Dagotto, Phys. Rev.
B {\bf 73}, 094434 (2006).

\bibitem{Mostovoy06} M. Mostovoy, Phys. Rev. Lett. {\bf 96},
067601 (2006).

\bibitem{Murakawa08} H. Murakawa, Y. Onose, K. Ohgushi, S.
Ishiwata, and Y. Tokura, J. Phys. Soc. Japan, {\bf 77}, 043709
(2008).

\bibitem{Tobbens01} D. M. Tobbens, N. St\"u$\ss$er, K. Knorr,
H. M. Mayer, and G. Lampert, Mater. Sci. Forum {\bf 378-381}, 288
(2001).

\bibitem{Rodriguez93} J. Rodriguez-Carvajal, Physica B {\bf 192},
55 (1993).

\bibitem{Fitch04} A. N. Fitch, J. Res. NIST {\bf 109}, 133 (2004).

\bibitem{Akimitsu78} J. Akimitsu, K. Siratori, G. Shirane, M. Iizumi,
and T. Watanabe, J. Phys. Soc. Japan, {bf 44}, 172 (1978).

\bibitem{Plumier75} R. Plumier, M. Lecomte, A. Mi\'edan-Gros, and
M. Sougi, Phys. Letters, {\bf 55A}, 239 (1975).

\bibitem{Krimmel06a} A. Krimmel, M. M\"ucksch, V. Tsurkan, M. M.
Koza, H. Mutka, C. Ritter, D. Sheptyakov, S. Horn, and A. Loidl,
Phys. Rev. B {\bf 73}, 014413 (2006).

\bibitem{Bergman07} D. Bergman, J. Alicea, E. Gull, S. Trebst, and
L. Balents, Nature Phys. {\bf 3}, 487 (2007).

\bibitem{Hamedoun86} M. Hamedoun, A. Wiedenmann, J. L. Dormann, M.
Nogues, and J. Rossat-Mignod, J. Phys. C {\bf 19}, 1783 (1986). M.
Hamedoun, A. Wiedenmann, J. L. Dormann, M. Nogues, and J.
Rossat-Mignot, J. Phys. C {\bf 19}, 1801 (1986).

\bibitem{Lee07} S.-H. Lee, G. Gasparovic, C. Borholm, M. Matsuda,
J.-H. Chung, Y. J. Kim, H. Ueda, G. Xu, P. Zschack, K. Kakurai, H.
Takagi, W. Ratcliff, T. H. Kim, and S.-W. Cheong, J. Phys.: Cond.
Matter {\bf 19}, 145259 (2007).



\bibitem{Hamilton58} W. C. Hamilton, Phys. Rev. {\bf 110}, 1050 (1958).


\end{thebibliography}
\end{document}